\documentclass[a4paper,12pt]{article}
\title{Bond Graph Representation of \\Chemical Reaction Networks}
\usepackage[noblocks]{authblk}
\author[1,2]{Peter J. Gawthrop\footnote{Corresponding
    author. \textbf{peter.gawthrop@unimelb.edu.au}}}
\author[1,2]{Edmund J. Crampin}
\affil[1]{
  Systems Biology Laboratory,
  Department of Biomedical Engineering,
  Melbourne School of Engineering,
  University of Melbourne,
  Victoria 3010, Australia.
   }
   \affil[2]{Systems Biology Laboratory,
     School of Mathematics and Statistics,
     University of
      Melbourne University of Melbourne, Victoria 3010}
\usepackage{a4wide}

\usepackage{cite}
\usepackage{amsmath,amssymb,amsfonts}
\usepackage{algorithmic}
\usepackage{graphicx}
\usepackage{textcomp}
\usepackage{xcolor}

\usepackage{siunitx}

\usepackage[numbers,compress]{natbib}
\usepackage[pagebackref,hidelinks]{hyperref}
\usepackage{url,doi}
\usepackage{makeidx} 

\makeindex  

\usepackage[section]{placeins}

\newcommand{\BG}[1]{\text{\sffamily\textbf{#1}}}

\newcommand{\C}{\BG{C }}

\newcommand{\Df}{\BG{Df }}

\newcommand{\one}{\BG{1 }}
\newcommand{\zero}{\BG{0 }}

\newcommand{\TF}{\BG{TF }}

\renewcommand{\Re}{\BG{Re }}

\usepackage{amsmath,amssymb,amscd,amstext,mathtools,extarrows,centernot}

\newcommand{\lb}{\left (}
\newcommand{\rb}{\right )}
\newcommand{\hp}{{\cdot}}         
\newcommand{\Exp}{\text{Exp}}
\newcommand{\Ln}{\text{Ln}}
\newcommand{\diag}{\text{diag}}
\newcommand{\where}{\text{where }}
\newcommand{\hence}{\text{hence }}
\newcommand{\an}{\text{and }}

\newcommand{\NN}{N}
\newcommand{\Ncd}{\NN^{cd}}
\newcommand{\Nf}{{\NN^f}}
\newcommand{\Nr}{{\NN^r}}

\newcommand{\ZZ}{Z}
\newcommand{\Zcd}{\ZZ^{cd}}
\newcommand{\DD}{D}
\newcommand{\Dcd}{\DD^{cd}}
\renewcommand{\Df}{{\DD^f}}
\newcommand{\Dr}{{\DD^r}}

\newcommand{\xx}{x}
\newcommand{\xs}{{\xx^\std}}

\newcommand{\XX}{X}
\newcommand{\Xs}{{\XX^\std}}

\newcommand{\Phif}{{\Phi^f}}
\newcommand{\Phir}{{\Phi^r}}
\newcommand{\Phic}{{\Phi^c}}
\newcommand{\phiN}{{\phi_N}}
\newcommand{\phis}{{\phi^\std}}

\newcommand{\KK}{K}
\newcommand{\Ks}{\KK^s}
\newcommand{\Kc}{\KK^c}
\newcommand{\Kv}{\KK^v}

\newcommand{\vc}{v^c}

\newcommand{\dx}{\dot{x}}

\newcommand{\dxf}{\dx^f}
\newcommand{\dxr}{\dx^r}

\newcommand{\vcf}{{\vc}^f}
\newcommand{\vcr}{{\vc}^r}

\usepackage{chemformula}

\newcommand{\std}{\oslash}

\usepackage{graphicx,subfigure,fancybox,color}

\newcommand{\Fig}[2]{
  \includegraphics[width=#2\columnwidth]{#1.pdf}
   \label{subfig:#1}
}

\newcommand{\SubFig}[3]{
  \subfigure[#2]{
    \includegraphics[width=#3\columnwidth]{#1.pdf}
    \label{subfig:#1}
  }
}

\begin{document}
\maketitle
\begin{abstract}
 The Bond Graph approach and the Chemical Reaction Network approach
 to modelling biomolecular systems developed independently. This paper
 brings together the two approaches by providing a bond graph
 interpretation of the chemical reaction network concept of
 complexes. Both closed and open systems are discussed.

 The method is illustrated using a simple enzyme-catalysed reaction
 and a trans-membrane transporter.
\end{abstract}
\newpage
\tableofcontents
\newpage
\section{Introduction}
\label{sec:introduction}
The bond graph method for modelling engineering systems
~\citep{Pay61,Cel91,GawSmi96,GawBev07,Bor10,KarMarRos12} was shown to
provide a thermodynamically consistent approach to modelling biomolecular
systems by ~\citet{OstPerKat71,OstPerKat73} and further developed by
\citet{GawCra14,GawCra16,GawCra17}. In this context, the relationship
between biomolecular systems and electrical circuit theory was
explored by \citet{OstPer74}.

In parallel with the seminal work of~\citet{OstPerKat71,OstPerKat73},
the mathematical foundations of chemical reaction networks (CRN) were
being laid by \citet{Fei72}, \citet{HorJac72} and \citet{FeiHor74}.
%
This approach to chemical reaction network theory was further
developed by \citet{Son06}, \citet{Ang09}, and
\citet{SchRaoJay13,SchRaoJay15,SchRaoJay16}. General results on
stability of both closed and open systems of chemical reactions have
been derived and applied to reveal dynamic features of complex
(bio)chemical networks \citep{ConFloRai07}, dissipation in noisy
chemical networks \citet{PolWacEsp15}, metabolic networks
\citep{IvaSchWei16} and multistability in interferon signalling
\citet{OteYorSte17}.

As an energy-based method, bond graphs are related to
\emph{port-Hamiltonians} \citep{GolSchBre03,VinBalGaw06,DonJun09}. A
port-Hamiltonian interpretation of CRNs has been given by
\citet{SchRaoJay13a} and this provides another link between CRNs and
bond graphs.

The formal concept of \emph{complexes} is essential to chemical
reaction network theory.  Complexes are the combination of chemical
species forming the substrate and products of the network reactions.
This paper links chemical reaction network theory to the bond graph
approach by incorporating the concept of complexes into bond graph
modelling of biomolecular systems.


\S~\ref{sec:basic-ideas} introduces the basic ideas of chemical
reaction networks from the stoichiometric point of view and
\S~\ref{sec:bg-appr-compl} gives a bond graph interpretation.
\S~\ref{sec:system-equations} shows how system equations can be
simplified using the complex approach.
\S~\ref{sec:open-systems-} discusses thermodynamically open systems.
%
\S~\ref{sec:conclusion} concludes the paper.

\section{The Stoichiometric Approach to Complexes}
\label{sec:basic-ideas}
The notion of complexes was defined by \citet{FeiHor74}: ``By the
complexes in a mechanism we mean the set of entities appearing before
or after arrows in that mechanism.'' where ``mechanism'' is a
generalisation of ``chemical reaction''.
This section introduces some
basic ideas relating to the use of complexes in describing chemical
reaction networks by means of the simple reaction network example
\begin{equation}
  \ch{A + E <>[ r1 ] C <>[ r2 ] B + E}\label{eq:ABE_reac}
\end{equation}
This example involves the four species \ch{A}, \ch{B}, \ch{C} and
\ch{E} and the two reactions \ch{r1} and \ch{r2}. It represents the
reaction \ch{A <> B} catalysed by the enzyme \ch{E} and with
intermediate complex \ch{C} \citep{KeeSne09}.
The substrate of reaction \ch{r1} is \ch{A + E} and the product is
\ch{C}; the substrate of reaction \ch{r2} is \ch{C} and the product is
\ch{B + E}.  Thus there are three complexes associated with this
reaction network: \ch{A + E}, \ch{C} and \ch{B + E}; \ch{C} forms not
only the right-hand side of reaction \ch{r1} but also the left-hand
side of reaction \ch{r2}.

The standard stoichiometric approach would be to define the species
state $x$ and reaction flow $v$ as:
\begin{xalignat}{2}
  x &=
  \begin{pmatrix}
    x_A \\ x_B \\ x_C \\ x_E
  \end{pmatrix}&
  v &=
  \begin{pmatrix}
    v_1 \\ v_2
  \end{pmatrix}
\end{xalignat}
where $v_1$ and $v_2$ are the flows associated with  reactions \ch{r1}
and \ch{r2} respectively.
The rate of change of species $\dx$ is then given in terms of the
stoichiometric matrix $N$ and reaction flow $v$ as%
\footnote{Although equation \eqref{eq:dx_ABEC} is linear,
as discussed in \S~\ref{sec:system-equations} the reaction flow $v$
is, in general, a \emph{nonlinear} function of the species state
$x$. In this particular case the expression for $v$ involves the
nonlinear terms $x_Ex_A$ and $x_Ex_B$.}:
\begin{xalignat}{2}\label{eq:dx_ABEC}
  \dx &= \NN v&
\text{where }
N &= 
\begin{pmatrix}
  -1 & 0 \\
  0 & 1 \\
  1 & -1 \\
  -1 & 1
\end{pmatrix}
\end{xalignat}

In contrast, the complex-based approach uses the complex flows $v^c$ as an
intermediate quantity. Thus define
\begin{equation}
  v^c = 
  \begin{pmatrix}
    v^c_1 \\ v^c_2 \\ v^c_3
  \end{pmatrix}
\end{equation}
where $v^c_1$, $v^c_2$ and $v^c_3$ are the flows associated with
complexes \ch{A + E}, \ch{C} and \ch{B + E} respectively.

The rate of change of species $\dx$ is given in terms of the 
matrix $Z$ and the complex flow $v^c$ as:
\begin{xalignat}{2}\label{eq:ABEC_Z}
  \dx &= Z v^c&
  \text{where }
  Z &=
  \begin{pmatrix}
    1&0&0\\
    0&0&1\\
    0&1&0\\
    1&0&1
  \end{pmatrix}
\end{xalignat}
and the complex flow $v^c$ is given in terms of the matrix $D$
and the reaction flow $v$ as:
\begin{xalignat}{2}
  v^c &= Dv \label{eq:ABEC_D}&
  \text{where }
  D &=
  \begin{pmatrix}
    -1&0\\
    1&-1\\
    0&1
  \end{pmatrix}
\end{xalignat}
If follows from Equations (\ref{eq:Z}) and (\ref{eq:D}) that $\dx = ZD
v$ and thus it follows from \eqref{eq:dx_ABEC} that
\begin{equation}
  \label{eq:ZD}
  N = ZD
\end{equation}

\begin{figure}[htbp]
  \centering
  \Fig{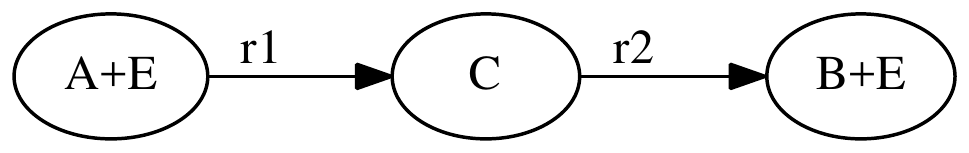}{0.7}
  \caption{Digraph corresponding to the $D$ matrix \eqref{eq:ABEC_D}
    for the system \ch{A + E <> C <> B + E}.  %
    The three complexes \ch{A + E}, \ch{C} and \ch{B + E} appear as
    nodes connected by a digraph with edges corresponding to the two
    reactions $r_1$ and $r_2$.  }
   \label{fig:ABEC_closed}
\end{figure}
The fundamental motivation for the complex-based approach is that
graph theory can be applied to the directed graph formed by taking the
complexes to be vertices and the reactions to be edges. In particular,
$D$ is the \emph{incidence matrix} of the graph and has the property
that each column of $D$ contains exactly one $1$ and exactly one $-1$;
the other elements being zero.
The corresponding digraph (plotted using \texttt{graphviz})
\citep{GanNor00}) appears in Figure~\ref{fig:ABEC_closed}.

Following \citet{GawCra14}, the stoichiometric matrix $\NN$ can be written as:
\begin{equation}\label{eq:NfNr}
  \NN = \Nr - \Nf
\end{equation}
where $\Nf$ and $\Nr$ connect the forward and reverse sides of the
reaction to \emph{species}. In a similar fashion, $\DD$ can be written
as:
\begin{equation}\label{eq:DfDr}
  \DD = \Dr - \Df
\end{equation}
where $\Df$ and $\Dr$ connect the forward and reverse sides of the
reaction to \emph{complexes}. $\Df$ and $\Dr$ can \emph{always} be
deduced from $\DD$ as $\Df$ and $\Dr$ correspond to the negative and
positive elements of $\DD$ respectively.

The columns of $\Nf$ correspond to the substrate complexes and that
the columns of $\Nr$ correspond to the product complexes. It follows
that the columns of both of these matrices contain all of the relevant
complexes, possibly repeated.
Hence $Z$ can be obtained as follows:
\begin{enumerate}
\item Create the matrix $Z_0$ from $\Nf$ and $\Nr$  and create the
  corresponding matrix $D_0$
  \begin{align}
    Z_0 &=
    \begin{pmatrix}
      \Nf & \vdots & \Nr
    \end{pmatrix}    \label{eq:Z0}\\
   D_0 &=
    \begin{pmatrix}
      -I_{n_V\times n_V}\\
      \hdots \\
      I_{n_V\times n_V}
    \end{pmatrix}    \label{eq:D0}
  \end{align}
It follows from Equation \eqref{eq:NfNr} that $Z_0D_0=\NN$.
\item Delete repeated columns of $Z_0$ to create $Z$ and sum the
  corresponding rows of $D_0$ to create $D$.
\end{enumerate}
Continuing the example of this section
\begin{xalignat}{2}
  \Nf &=
  \begin{pmatrix}
    1&0\\
    0&0\\
    0&1\\
    1&0
  \end{pmatrix}&
  \Nr &=
  \begin{pmatrix}
    0&0\\
    0&1\\
    1&0\\
    0&1
  \end{pmatrix}
\end{xalignat}
and so
\begin{xalignat}{2}
  Z_0 &=
  \begin{pmatrix}
    1&0&0&0\\
    0&0&0&1\\
    0&1&1&0\\
    1&0&0&1
  \end{pmatrix}&
  D_0 &=
  \begin{pmatrix}
    -1&0\\
    0&-1\\
    1&0\\
    0&1
  \end{pmatrix}
\end{xalignat}
As columns two and three are identical, column three of $Z_0$ is
deleted to give $Z$~\eqref{eq:ABEC_Z}, and rows two and three of $D_0$ are
merged to give $D$~\eqref{eq:ABEC_D}.





\section{The Bond Graph Approach to Complexes}
\label{sec:bg-appr-compl}
\begin{figure}[htbp]
  \centering
  \SubFig{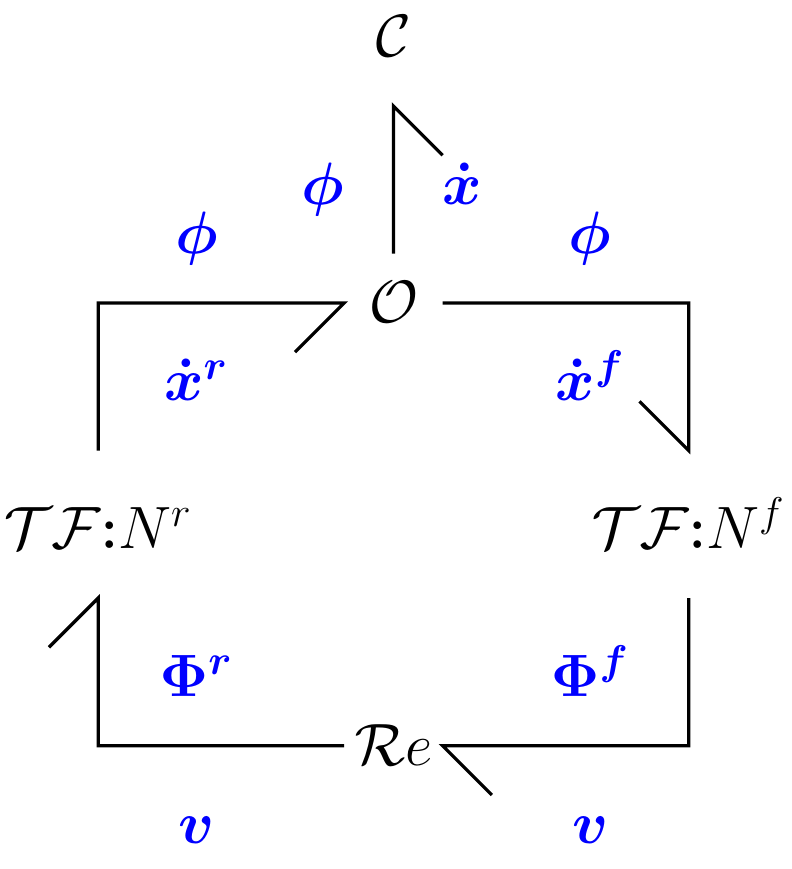}{Standard approach}{0.45}
  \SubFig{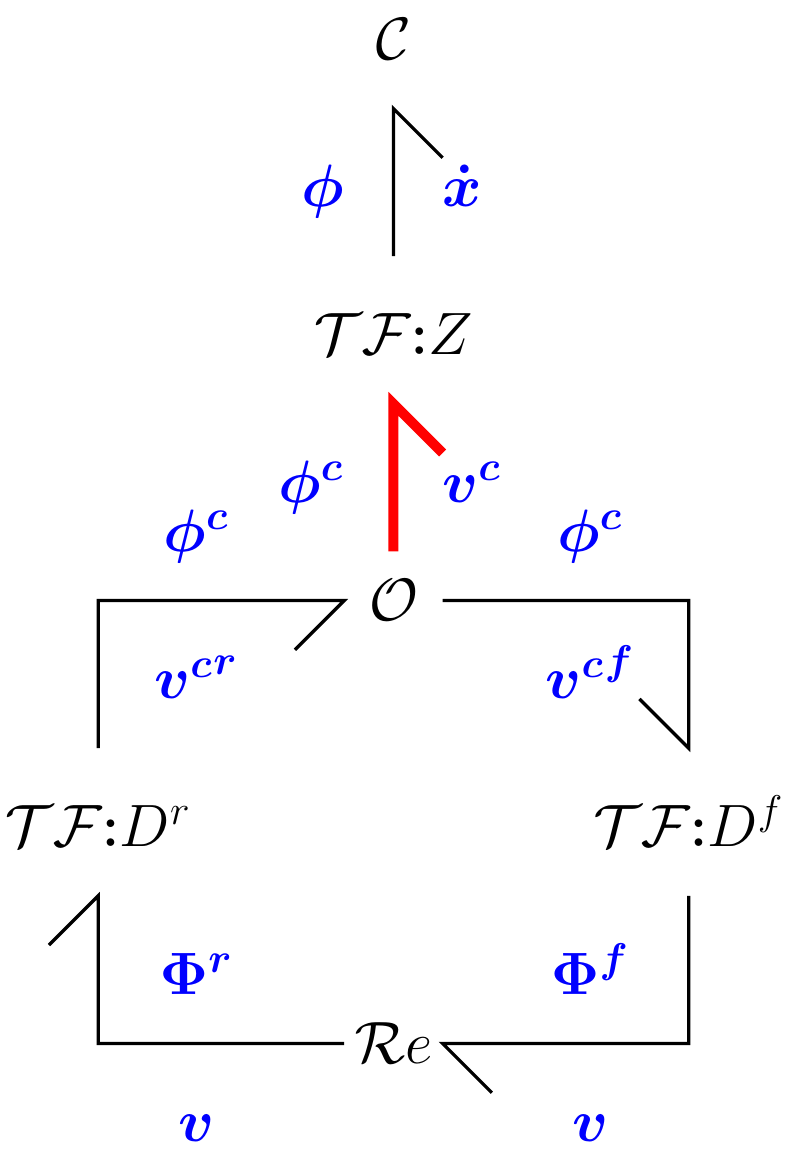}{Complex-based approach}{0.45}
  \caption{A Bond Graph Approach to Complexes.
    (a) The standard approach given by
    \citet{GawCra14,GawCra16,GawCra17}.
    The bond symbols $\rightharpoondown$ correspond
    to \emph{vectors} of bonds; $\mathcal{C}$, $\mathcal{R}e$ and
    $\mathcal{O}$ correspond to arrays of \C, \Re and \zero
    components; the 
    two $\mathcal{TF}$ components represent the intervening
    junction structure comprising bonds, \zero and \one junctions and
    \TF components. $N^f$ and $N^r$ are the forward and reverse
    stoichiometric matrices.
    (b) The complex based approach.
    $\mathcal{TF}{:}Z$ represents the junction structure connecting
    complexes and species where $Z$ appears in Equation \eqref{eq:Z}
    and  $\mathcal{TF}{:}\Df$ $\mathcal{TF}{:}\Dr $represent the junction structure connecting
    reactions and complexes where $\Df$ and $\Dr$ appear in Equations
    \eqref{eq:D} and \eqref{eq:DfDr}.
  }
\end{figure}

Figure \ref{subfig:Closed_bg} shows the approach used by
\citet{GawCra14,GawCra16,GawCra17} to represent \emph{closed
  systems}. However, following the approach of \citet{Gaw17a}, the
\emph{Faraday-equivalent potential} $\phi$, with units of \si{V}, is
used in place of chemical potential $\mu$ with units of
\si{J.mol^{-1}}.
$\mathcal{C}$ represents the $n_s$ \C components representing the
chemical species; $\phi$ is the vector of the chemical potentials, and
$\dx$ the corresponding flow rates. $\mathcal{R}e$ represents the
$n_r$ \Re components representing the chemical reactions with forward
and reverse potential $\Phif$ and $\Phir$ and flow rate
$v$. $\mathcal{TF}$:$N^r$ and $\mathcal{TF}$:$N^r$ represent the bond
graph \emph{transformers} encapsulating the system stoichiometry.
A key feature of transformers is that they  relate both the efforts and
flows on the corresponding bonds whilst conserving energy
\citep{GawCra14,GawCra16,GawCra17}. Thus, with reference to
Figure~\ref{subfig:Closed_bg}
\begin{align}
  \dx &= \dxr-\dxf = \Nr v - \Nf v = N v \label{eq:N}\\
  \Phi &= \Phi^f - \Phi^r = \Nf^T \phi - \Nr^T \phi  \notag\\
      &= -N^T \phi \label{eq:NT}
\end{align}

In contrast, Figure \ref{subfig:cClosed_bg} shows the complex-based
approach used here. $\mathcal{TF}$:$Z$ represents the bond graph
\emph{transformer} relating the $n_s$ species to the $n_c$
\emph{complexes} which then become the reaction forward complex
(substrates) via $\mathcal{TF}$:$D^f$ and the reaction reverse complex
(products) via $\mathcal{TF}$:$D^r$.
With reference to Figure~\ref{subfig:cClosed_bg}, the transformer
equations become:
\begin{align} \dx &= Z v^c\label{eq:Z}\\ \phi^c &= Z^T
\phi\label{eq:ZT}\\ \vc &= \vcr-\vcf = \Dr v - \Df v = D
v \label{eq:D}\\ \Phi &= \Phi^f - \Phi^r = \Df^T \phi^c - \Dr^T \phi^c
\notag\\ &= -D^T \phi^c \label{eq:DT}
\end{align}

    %

\subsection{Example: \ch{A + E <> C <> B + E}}
\begin{figure}[htbp]
  \centering
  \SubFig{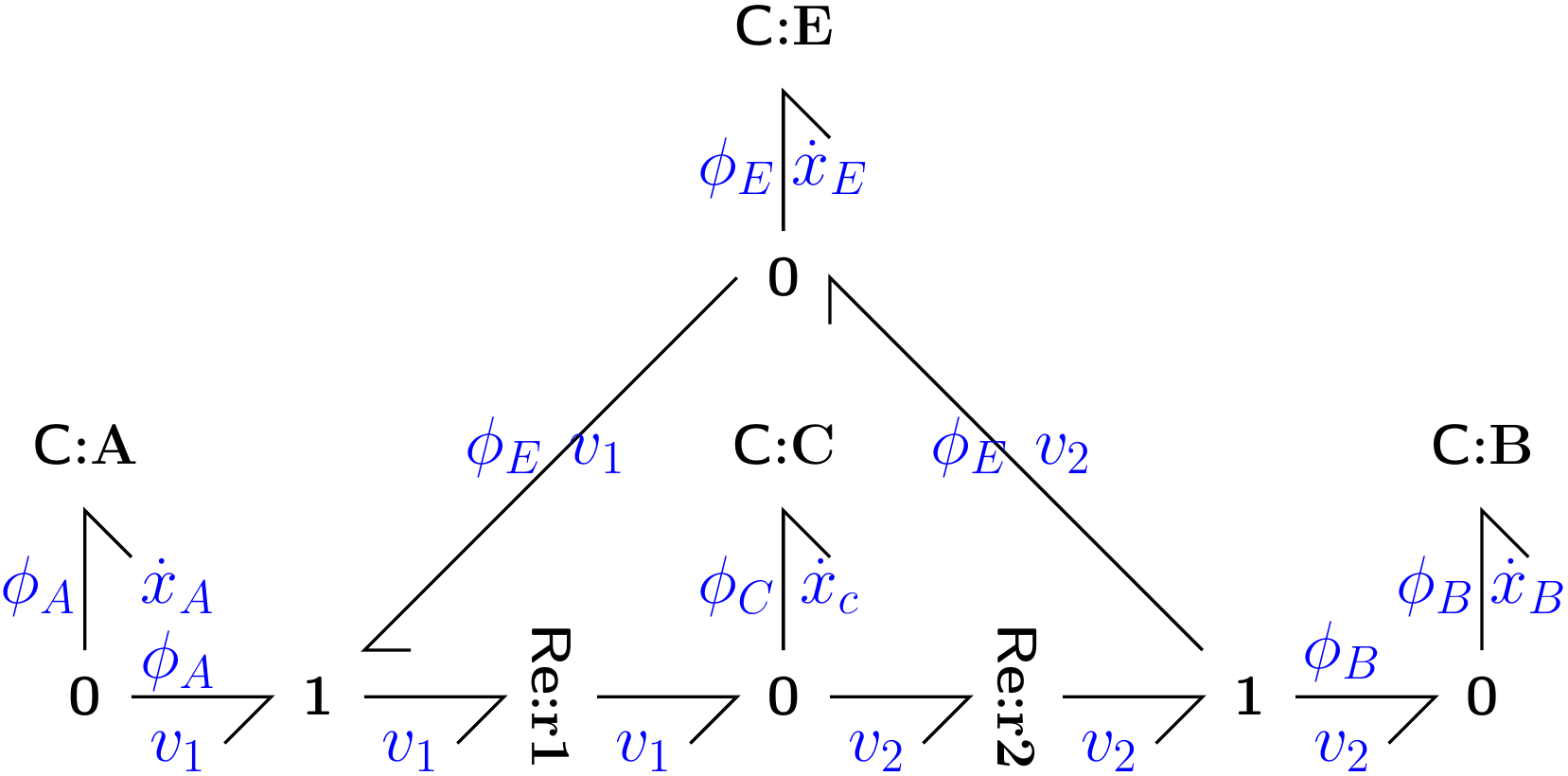}{Standard approach}{0.4}
  \SubFig{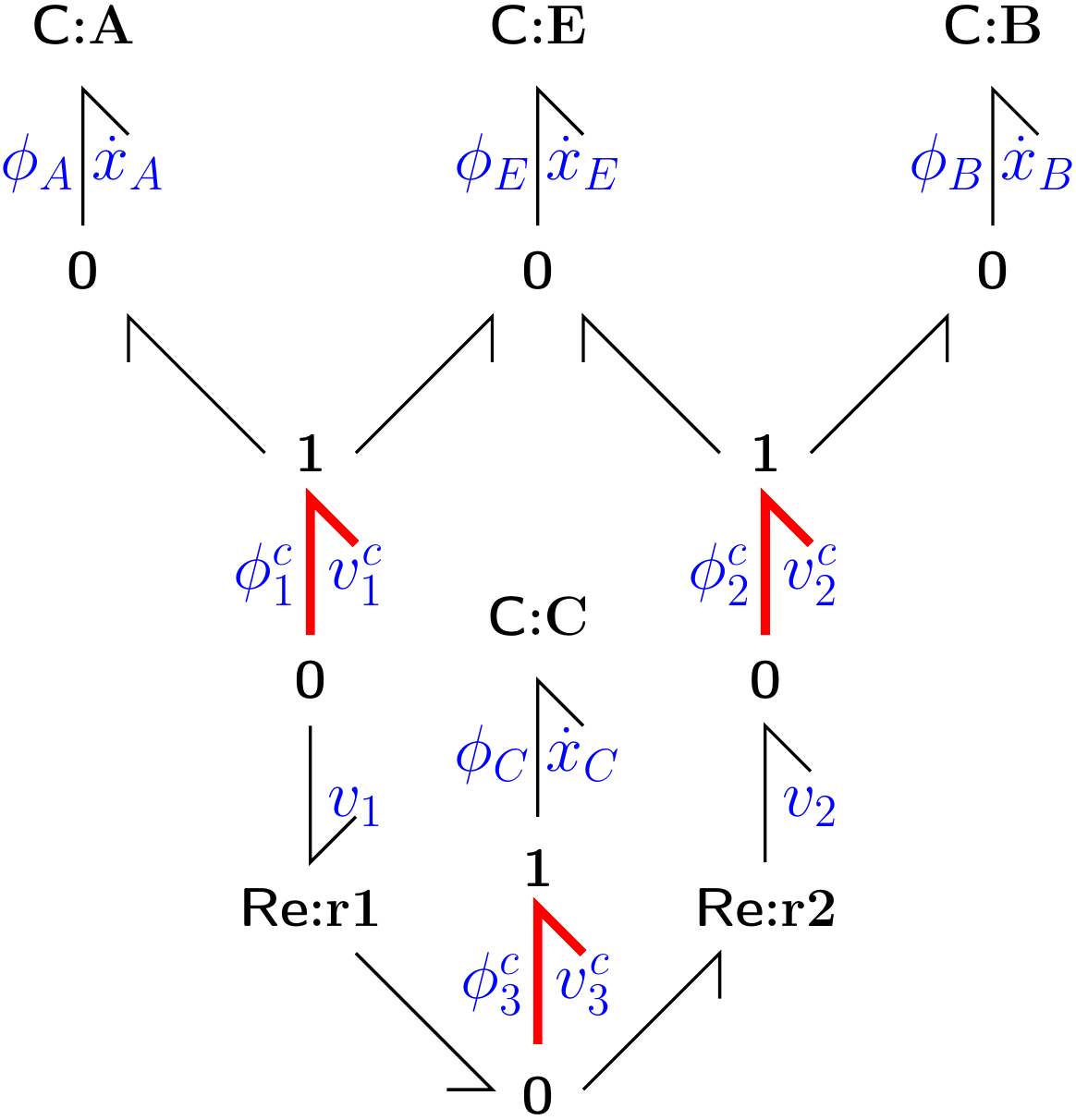}{Complex-based approach}{0.3}
  \caption{Example: \ch{A + E <> C <> B + E}.
    (a) A bond graph without explicit representation of complexes.
    (b) The complex covariables correspond to the three highlighted
    bonds.
    The junction structure connecting the three highlighted
    bonds to the species corresponds to $\mathcal{TF}{:}Z$ of Figure
    \ref{subfig:cClosed_bg}
    and the junction structure connecting the reaction \Re components
    to the three highlighted
    bonds  corresponds to $\mathcal{TF}{:}\Df$ and
    $\mathcal{TF}{:}\Dr$ of Figure
    \ref{subfig:cClosed_bg}. Bonds pointing \emph{into} the \Re
    components correspond to $\mathcal{TF}{:}\Df$, those pointing
    \emph{away from} the \Re
    components correspond to $\mathcal{TF}{:}\Dr$.
}
\end{figure}

Figure \ref{subfig:ABEC_abg} gives the standard bond graph for the
reaction \ch{A + E <> C <> B + E} corresponding to the general Figure
\ref{subfig:Closed_bg} and Figure \ref{subfig:cABEC_abg} gives the
complex-based bond graph corresponding to the general Figure
\ref{subfig:cClosed_bg}. The three bonds corresponding to the three
complex efforts $\Phic$ and flows $\vc$ are highlighted. 




\section{System equations}\label{sec:system-equations}
This section derives the main properties of the CRN modelled by a bond
graph including the complex concept. 
The notation and concepts of \citet[\S~2]{SchRaoJay16} are used and
reversible reactions are used at the outset. 
Following the notation of \citet{SchRaoJay16}, $\hp$,
$\frac{\xx}{\xs}$, $\Ln$ and $\Exp$ denote elementwise multiplication,
division, natural logarithm and exponentiation of column vectors. In
particular for two column vectors $x$ and $y$:
\begin{xalignat}{2}\label{eq:hp}
  x \hp y &= \diag  (x)  y &
  \frac{x}{y} = \lb \diag (y) \rb^{-1} x
\end{xalignat}

The basic equation for the potential of species expressed as the
Faraday-equivalent potential \citep{Gaw17a} is
\begin{align}
  \phi &= \phis + \phiN\Ln \frac{\xx}{\xs}\label{eq:phi}\\
  \where
  \phiN &= \frac{RT}{F}  \approx \SI{26}{mV}
\end{align}

Alternatively, \eqref{eq:phi} can be rewritten as
\begin{align}
  \phi &= \phiN\Ln \lb \Ks \hp \xx \rb\label{eq:phiK}\\
  \where
  \Ks &= \frac{\exp \frac{\phis}{\phiN}}{\xs}
\end{align}

\subsection{Properties of the complexes}
The basic bond graph notion of transformers as expressed in Figure
\ref{subfig:cClosed_bg} means that the potential of the complexes can be
expressed as:
\begin{align}
  \Phic &= Z^T\phi\\
\hence
  \Phic &=  \Phic^\std + \phiN \Ln \frac{X}{X^\std}\\
  \text{where }
  \Phic^\std &= Z^T \phi^\std\\
  \XX &= \Exp \lb Z^T  \Ln \xx \rb  = \prod_{j=1}^{n_s} \xx_j^{z_{ji}}\\
  \text{and }
  \Xs &= \Exp \lb Z^T  \Ln \xs \rb  = \prod_{j=1}^{n_s} {\xs_j}^{z_{ji}}
\end{align}
Alternatively, using Equation \eqref{eq:phiK}
\begin{align}
  \Phic &= \phiN\Ln \lb \Kc \hp \XX \rb \label{eq:PhiK}\\
  \where
  \Kc &= \Exp \lb Z^T  \Ln \Ks \rb  = \prod_{j=1}^{n_s} {k^s_j}^{z_{ji}}\label{eq:Kc}
\end{align}
Using Equation \eqref{eq:hp}, Equation \eqref{eq:PhiK} can also be
written as
\begin{equation}\label{eq:PhiK_alt}
  \Phic  = \phiN\Ln \lb \diag \Kc  \XX \rb\
\end{equation}

\subsection{Mass-action Kinetics}
Mass action kinetics correspond to the Marcelin-de Donder
formula~\citep{Rys58,OstPerKat73,GawCra14}:
\begin{align}
  v &= \kappa \hp \lb \Exp \frac{\Phif}{\phiN} 
      - \Exp \frac{\Phir}{\phiN}\rb \label{eq:v_0}\\
  \where
  \Phif &= \Nf^T\phi = \Df^T \Phic\\
  \an
  \Phir &= \Nr^T\phi = \Dr^T \Phic
\end{align}
Using Equation \eqref{eq:PhiK}, \eqref{eq:v_0} becomes:
\begin{align}
  v &= \kappa \hp \lb \Exp \Df^T \frac{\Phic}{\phiN} 
      - \Exp \Dr^T\frac{\Phic}{\phiN}\rb \label{eq:v_1}
\end{align}
Because the matrices $\Df^T$ and $\Dr^T$ are simply selecting the
appropriate complexes for each reaction, each row has exactly one unit
element and the rest zero. Hence Equation \eqref{eq:v_1} becomes:
\begin{align}
  v &= \kappa \hp \lb \Df^T  \Exp \frac{\Phic}{\phiN} 
      -  \Dr^T\Exp\frac{\Phic}{\phiN}\rb \notag\\
    &=  -\kappa \hp \DD^T \Exp\frac{\Phic}{\phiN} \label{eq:v_2}
\end{align}
Using Equation \eqref{eq:PhiK_alt} Equation \eqref{eq:v_2} becomes:
\begin{align}
  v & = \Kv \XX = \Kv \Exp \lb Z^T  \Ln \xx \rb\label{eq:v=KvXX}\\
  \where
  \Kv &= -\kappa \hp \lb \DD^T \diag{\Kc}\rb\label{eq:Kv}
\end{align}

Hence the system state equation for mass action kinetics is:
\begin{align}\label{eq:dx_MA_closed}
  \dx &= \ZZ \DD \Kv  \XX \notag\\ 
      &= \NN \Kv  \XX \notag\\ 
      &= \NN \Kv \Exp \lb Z^T  \Ln \xx \rb
\end{align}
This is essentially Equation (4) of \citet{SchRaoJay16}.
Note that the term $\Exp \lb Z^T  \Ln \xx\rb$ appearing in equations
\eqref{eq:v=KvXX} and \eqref{eq:dx_MA_closed}
is, in general, nonlinear. As will be seen in the following section,
this term leads to \emph{products} of species states.

\subsection{Example: \ch{A + E <> C <> B + E} (continued)}

Substituting the numerical values from the example of
\S~\ref{sec:basic-ideas} into Equation \eqref{eq:Kc}:
\begin{align}
  \Kc &= \Exp \lb Z^T  \Ln \Ks \rb \notag\\
      &= \Exp
        \begin{pmatrix}
          1&0&0&1\\
          0&0&1&0\\
          0&1&0&1
        \end{pmatrix}
                 \begin{pmatrix}
                   \ln \Ks_A\\
                   \ln \Ks_B\\
                   \ln \Ks_C\\
                   \ln \Ks_E\\
                 \end{pmatrix} \notag\\
      &= \Exp
        \begin{pmatrix}
          \ln \Ks_A + \ln \Ks_E\\
          \ln \Ks_C\\
          \ln \Ks_B + \ln \Ks_E\\
        \end{pmatrix} \notag\\
      &=  \begin{pmatrix}
        \Ks_A \Ks_E \\\Ks_C\\ \Ks_B \Ks_E
      \end{pmatrix}
\end{align}
Similarly:
\begin{align}
    \XX &=
  \begin{pmatrix}
    x_A x_E\\x_C\\x_B x_E
  \end{pmatrix}
\end{align}

Substituting the numerical values from the example of
\S~\ref{sec:basic-ideas} into Equation \eqref{eq:Kv}
\begin{align}
\Kv &= -\kappa \hp \lb \DD^T \diag{\Kc}\rb  \notag\\
    &= -\kappa \hp 
      \begin{pmatrix}
        -1&1&0\\
        0&-1&1
      \end{pmatrix}
              \diag{\Kc} \notag \\
    &= 
  \begin{pmatrix}
    \kappa_1 \Ks_A \Ks_E  & -\kappa_1 \Ks_C & 0\\
    0 & \kappa_2 \Ks_C & -\kappa_2 \Ks_B \Ks_E \\
  \end{pmatrix}
\end{align}
Hence, using \eqref{eq:v=KvXX}
\begin{align}
  v & = \Kv \XX =
      \begin{pmatrix}
        \kappa_1 \lb \Ks_A \Ks_E x_A x_E -   \Ks_C x_c\rb\\
        \kappa_2  \lb \Ks_C x_c -  \Ks_B \Ks_E x_B x_E \rb
      \end{pmatrix}
\end{align}

\subsection{Example: Transporter}\label{sec:example:-transporter}
\begin{figure}[htbp]
  \centering
  \SubFig{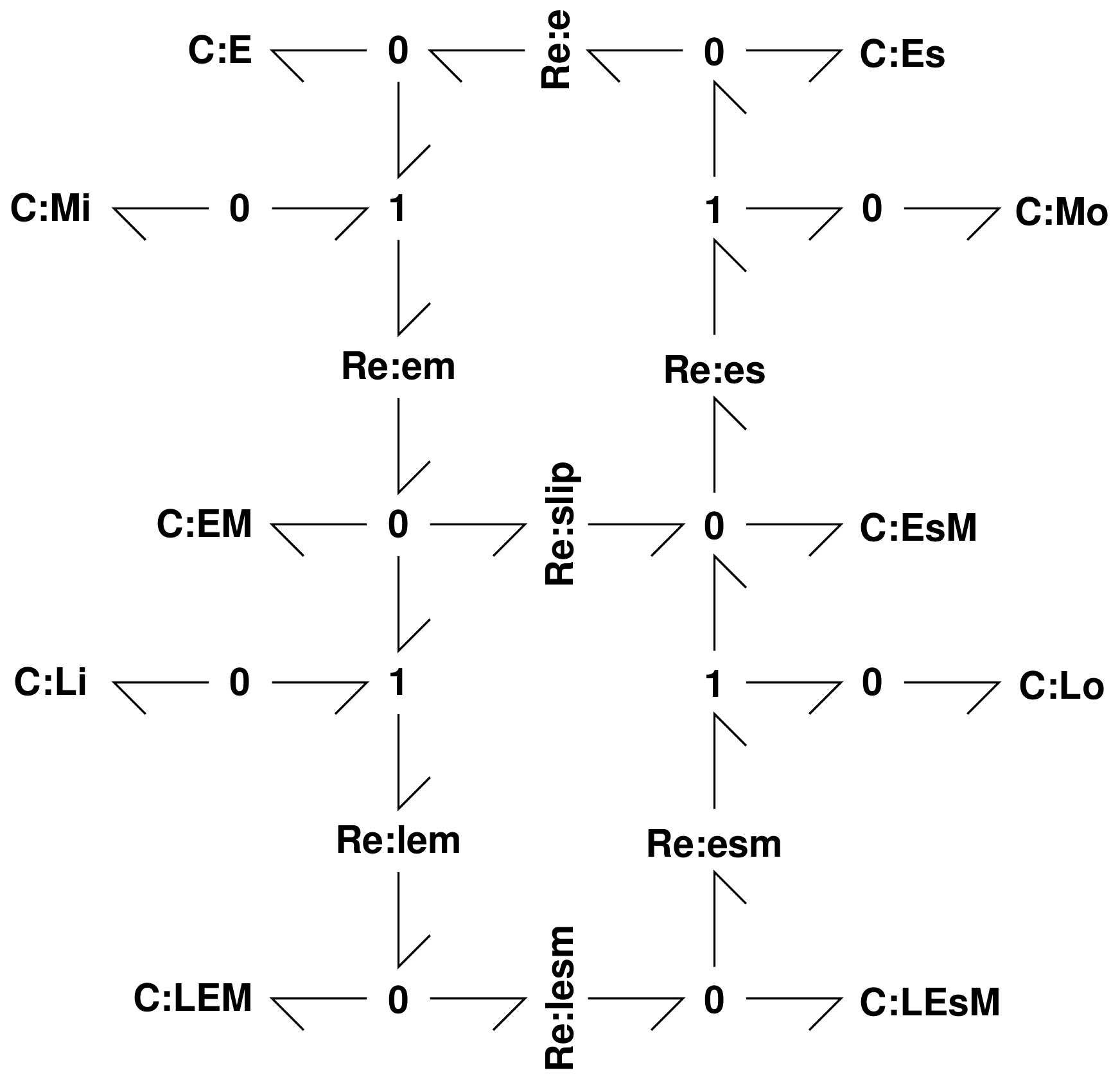}{Standard approach}{0.7}
  \SubFig{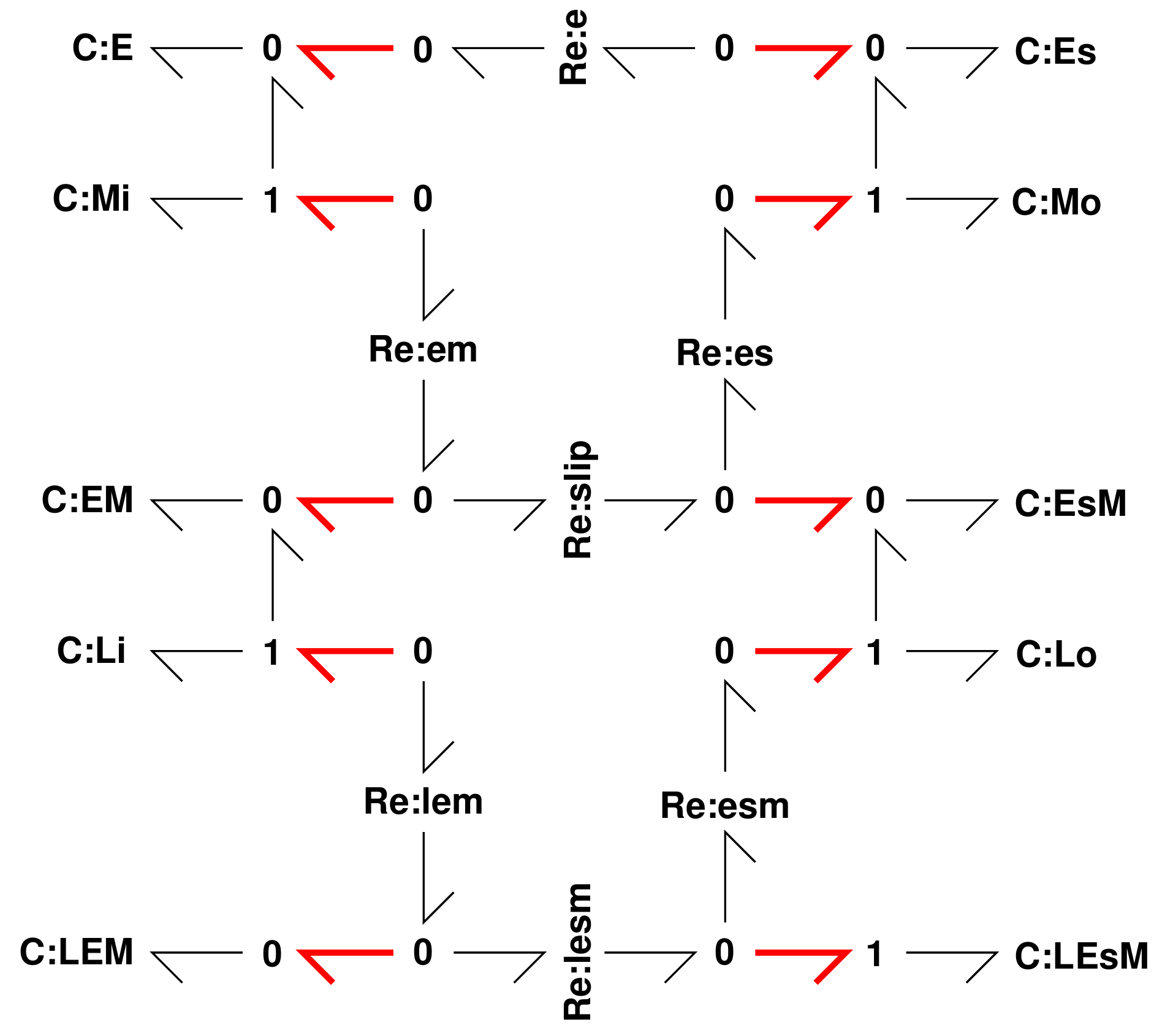}{Complex-based approach}{0.7}
  \caption{Example: Transporter \citep{Hil89}.
    (a) The bond graph without explicit representation of complexes
    \citep{GawCra17}.
    (b) The ten complexes correspond to the ten highlighted
    bonds. 
    The junction structure connecting the ten highlighted
    bonds to the species corresponds to $\mathcal{TF}{:}Z$ of Figure
    \ref{subfig:cClosed_bg}
    and the junction structure connecting the reaction \Re components
    to the ten highlighted
    bonds  corresponds to $\mathcal{TF}{:}\Df$ and
    $\mathcal{TF}{:}\Dr$ of Figure
    \ref{subfig:cClosed_bg}.
  }
\end{figure}
The seminal book ``Free energy transduction and biochemical cycle
kinetics'' of \citet{Hil89} contains an example of a membrane
transporter which is discussed in detail by \citet{GawCra17}. The bond
graph is given in Figure \ref{subfig:Hills_abg} and the bond graph redrawn
to expose the complexes is given in Figure \ref{subfig:cHills_abg};
the ten bonds corresponding to the ten complex efforts $\Phic$ and
flows $\vc$ are highlighted.

The ten complexes are: 
\ch{ E}, 
\ch{ Mi + E},
\ch{ EM},
\ch{ Li + EM},
\ch{ LEM},
\ch{ LEsM},
\ch{ Lo + EsM},
\ch{ EsM},
\ch{ Mo + Es} and 
\ch{ Es}.
They are connected by the seven reactions
\ch{ em},
\ch{ lem},
\ch{ lesm},
\ch{ esm},
\ch{ es},
\ch{ e} and
\ch{ slip}.
The corresponding digraph (plotted using \texttt{graphviz})
\citep{GanNor00}) appears in Figure~\ref{subfig:Hills_closed_pos}.

\subsection{Michaelis-Menten Kinetics}
Enzyme-catalysed reactions such as \eqref{eq:ABE_reac},
\S~\ref{sec:basic-ideas} can be approximated to give Michaelis-Menten
kinetics. In particular, in the bond graph  context, \citet{GawCra14}
show that the two reactions of \eqref{eq:ABE_reac}, generalised to allow
multiple products and reactants, can be replaced by a single reaction
with equivalent rate-constant $\kappa_e$ given in terms of the rate
constants $\kappa_1$ and $\kappa_2$ of the reactions $r_1$ and $r_2$ as
\label{sec:mich-ment-kinet}
\begin{align}
 \kappa_e &= e_0 \frac{\bar{\kappa} K_c}{k_m + \sigma_v} \\
  \text{ where } k_m &= \frac{K_c}{K_e} \label{eq:kappa_ecr}\\
  \bar{\kappa} &= \frac{\kappa_1 \kappa_2}{{\kappa_1+\kappa_2}}\\
  \text{and }
   \sigma_v &=
 \begin{cases}
   \frac{\exp{\frac{\Phif}{RT}} + \exp{\frac{\Phir}{RT}}}{2} & \kappa_1 = \kappa_2\\
   \exp{\frac{\Phif}{RT}} & \kappa_1 \gg \kappa_2
 \end{cases}
\end{align}
where $\Phif$ and $\Phir$ are the overall  forward
and reverse reaction potentials and $e_0$ is the total amount of
enzyme both free and bound to \ch{C}. In particular, in the case of the
reactions of \eqref{eq:ABE_reac}:
\begin{align}
  \sigma_v &=
             \begin{cases}
               \frac{K_A x_A + K_B x_B}{2} & \kappa_1 = \kappa_2\\
               K_A x_A & \kappa_1 \gg \kappa_2
             \end{cases}
\end{align}

When dealing with networks of enzyme catalysed reactions such as
\eqref{eq:ABE_reac} there are two choices: either explicitly model the
intermediate species \ch{C} and use two reactions with \emph{constant}
values of $\kappa$ or use a single reaction approximation without
intermediate species \ch{C} and an equivalent rate-constant
$\kappa_e$ which is a function of the species states $\xx$.

However, as discussed by \citet{Gun14}, this approximation should be
used with care to avoid violating the fundamental laws of
thermodynamics. For example, when modelling networks such as the
mitogen-activated protein kinase (MAPK) cascade where enzymes compete
and are themselves reaction products, it has been argued \citep[\S
9.5]{Voi13} that the mass-action approach is preferable. This
discussed in detail by \citet{GawCra16}.

Nevertheless, the bond graph representation of chemical reaction
networks used in this paper, although developed in the context of
mass-action kinetics, can equally be applied to systems approximated
using Michaelis-Menten kinetics. The difference is that the rate
constant $\kappa$ is replaced by an expression $\kappa_e(\xx)$
dependent on species states $\xx$.





\section{Open systems \& Chemostats}\label{sec:open-systems-}
There are a number of ways of converting closed systems to open
systems whilst retaining the basic closed system formulation.
\citet{HorJac72} use the concept of a \emph{zero complex} to act as a
generalised source and sink of chemical species and this idea is
followed up by \citet{SchRaoJay16}.
\citet{PolEsp14} use the concept of a \emph{chemostat} to act as a
source and sink of chemical species at fixed concentration and this
idea is followed up by \citet{GawCra16}. 
The chemostat has three interpretations:
\begin{enumerate}
\item one or more species is fixed to give a constant concentration
  \citep{GawCurCra15}; this implies that an appropriate external
  flow is applied to balance the internal flow of the species.
\item an ideal feedback controller is applied to species to be fixed
  with setpoint as the fixed concentration and control signal an
  external flow.
\item as a \C component with a fixed state.
\end{enumerate}
The chemostat approach is
used here.

As discussed by \citet{GawCra16}, for each species set to be a
chemostat, the corresponding row in the stoichiometric matrix $\NN$ is
replaced by a zero vector to form the \emph{chemodynamic}
stoichiometric matrix $\Ncd$. Using the same motivation as that
leading to equation \eqref{eq:ZD}, $\Ncd$ is written as:
\begin{equation}
  \label{eq:ZD_cd}
  N^{cd} = Z^{cd}D^{cd}
\end{equation}
In this case, the closed-system equations \eqref{eq:Z}--~\eqref{eq:DT}
are replaced by
\begin{align}
  \dx &= Z^{cd} v^c\label{eq:oZ}\\
  \phi^c &= Z^T \phi\label{eq:oZT}\\
  \vc &= \vcr-\vcf = D^{cd} v \label{eq:oD}\\
  \Phi &= \Phi^f - \Phi^r = \Df^T \phi^c - \Dr^T \phi^c  \notag\\
      &= -D^T \phi^c \label{eq:oDT}
\end{align}
Note that it is the \emph{flow} equations \eqref{eq:oZ} and
\eqref{eq:oD} that are changed; the potential equations \eqref{eq:oZT}
and \eqref{eq:oDT} remain the same as those for the closed system
\eqref{eq:ZT} and \eqref{eq:DT}. In particular, some complexes
associated with $Z$ and $D$, and thus the potential equations
\eqref{eq:oZT} and \eqref{eq:oDT} are not associated with $\Zcd$ and
$\Dcd$, and thus the flow equations \eqref{eq:oZ} and \eqref{eq:oD}.
Hence the digraph associated with $\Dcd$ does not necessarily contain
all of the complex nodes associated with $\DD$.

\subsection{Example: \ch{A + E <> C <> B + E} (continued)}
\begin{figure}[htbp]
  \centering
  \Fig{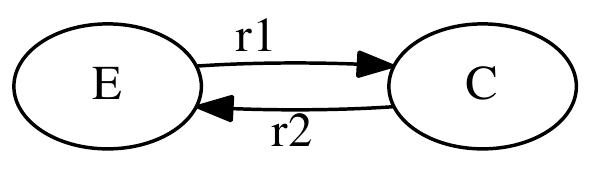}{0.4}
  \caption{Digraph corresponding to the $D^{cd}$ matrix
    \eqref{D_cd_simple} for the system \ch{A + E <> C <> B + E} of
    \S~\ref{sec:basic-ideas} and \S~\ref{sec:bg-appr-compl}.
    Compared to Figure \ref{fig:ABEC_closed}, setting the species
    \ch{A} and \ch{B} to be chemostats reduces the number of complexes
    to two and the digraph is cyclic.  }
  \label{fig:ABEC_open}
\end{figure}
In the case of the system \ch{A + E <>[ r1 ] C <>[ r2 ] B + E} and
choosing the two species \ch{A} and \ch{B} to be chemostats, equation
\eqref{eq:dx_ABEC} is replaced by:
\begin{xalignat}{2}\label{eq:dx_ABEC_cd}
  \dx &= \Ncd v&
\text{where }
N^{cd} &= 
\begin{pmatrix}
  0 & 0 \\
  0 & 0 \\
  1 & -1 \\
  -1 & 1
\end{pmatrix}
\end{xalignat}
Thus the two chemostats have constant state $x_A$ and $x_B$.
The decomposition of Equation \eqref{eq:ZD_cd} gives:
\begin{xalignat}{2}
  Z^{cd} &= 
  \begin{pmatrix}
    0 & 0 \\
    0 & 0 \\
    0 & 1 \\
    1 & 0
  \end{pmatrix}
  D^{cd} &= 
  \begin{pmatrix}
    -1 & 1 \\
    1 & -1
\end{pmatrix}\label{D_cd_simple}
\end{xalignat}
The digraph corresponding to $\Dcd$ is given in
Figure~\ref{fig:ABEC_open}; this corresponds to the flow equations
\eqref{eq:oZ} and \eqref{eq:oD}. On the other hand, the digraph
corresponding to $\DD$ is given in Figure~\ref{fig:ABEC_closed}; this
corresponds to the potential equations \eqref{eq:oZT} and
\eqref{eq:oDT}.
Thus the cyclic flow associated with the digraph of
Figure~\ref{fig:ABEC_open} is driven by the potentials associated with
the digraph of Figure~\ref{fig:ABEC_closed}.
\subsection{Example: Transporter (continued)}
\begin{figure}[htbp]
  \centering
  \SubFig{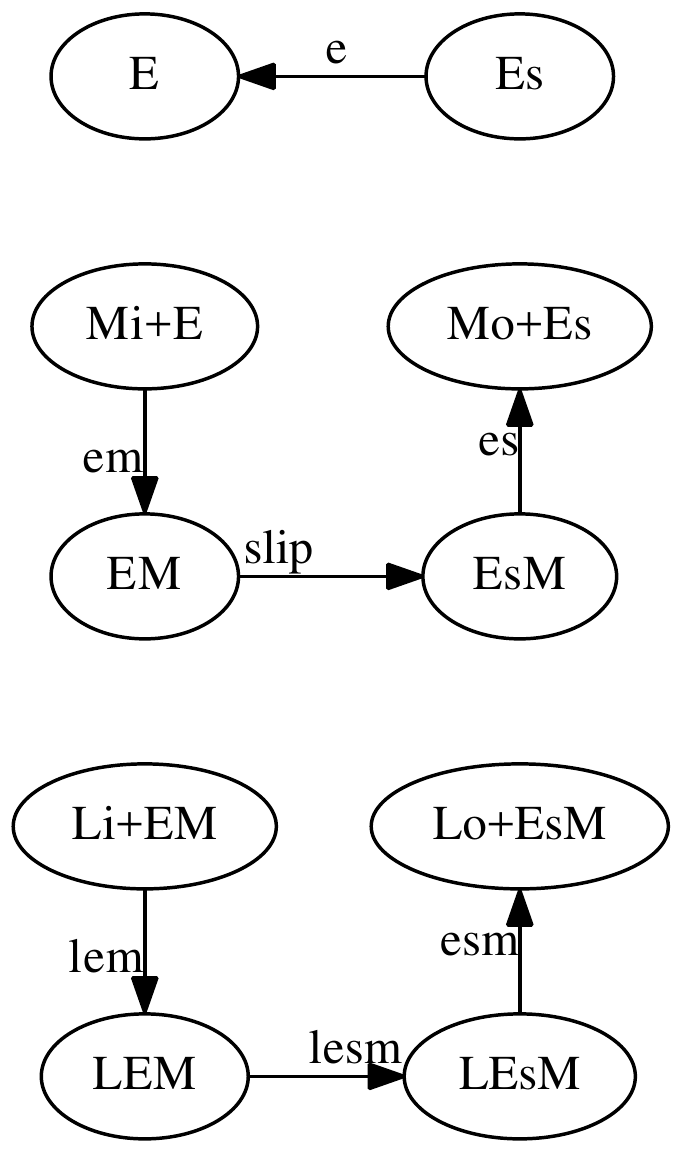}{Closed}{0.45}
  \SubFig{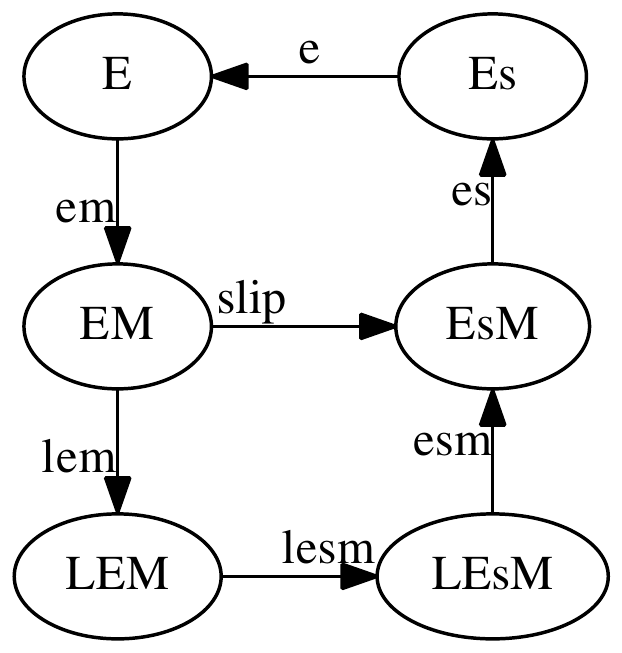}{Open}{0.45}
  \caption{Digraphs corresponding to the $D$
    matrix \eqref{eq:D} for the closed and open systems for the
    transporter system.
    (a) The ten nodes corresponding to the ten complexes are connected
    by three disjoint linear graphs.
    (b) The four chemostats reduce the number of complexes to six and
    the corresponding six nodes are connected by a cyclic digraph.
  }
  \label{fig:transporter_graphs}
\end{figure}
The closed system digraph, corresponding to $D$ and the potential
equations of the open system, is given in
Figure~\ref{subfig:Hills_closed_pos}.

As discussed by \citep{GawCra17}, the open system is created by choosing
the four species:
\ch{ Li},
\ch{ Lo},
\ch{ Mi} and 
\ch{ Mo} to be chemostats.
The flow digraph with incidence matrix $\Dcd$ of
Figure~\ref{subfig:Hills_open_pos} has six nodes corresponding to the
complexes: \ch{ E}, \ch{ EM}, \ch{ LEM}, \ch{ LEsM}, \ch{ EsM} and
\ch{ Es}.  This digraph still has the seven connecting reactions
listed in \S~\ref{sec:example:-transporter}.

The cyclic flow associated with the digraph of
Figure~\ref{subfig:Hills_open_pos} is driven by the potentials associated with the digraph of
Figure~\ref{subfig:Hills_closed_pos}.

\section{Conclusion}
\label{sec:conclusion}
The complex approach to modelling chemical reaction networks as
introduced by \citet{Fei72}, \citet{HorJac72} and \citet{FeiHor74} and
expanded by \citet{SchRaoJay13,SchRaoJay13a,SchRaoJay15,SchRaoJay16} has been given
a bond graph interpretation thus enabling results from the complex
approach to be applied to the bond graph approach and \emph{vice
  versa}. 
In particular, the decomposition of the stoichiometric
matrix $\NN$ into the complex composition matrix \citep{SchRaoJay16}
$Z$ and the complex graph incidence matrix $D$ (where $N=ZD$) is given
a bond graph interpretation.

The approach is developed for closed systems, but extended to open
systems via the previously developed notion of chemostats
\citep{PolEsp14,GawCra16}. The corresponding chemodynamic
stoichiometric matrix $\Ncd$ \citep{GawCra16} is decomposed into the
chemodynamic complex composition matrix $\Zcd$ and the chemodynamic
complex graph incidence matrix $\Dcd$ (where $\Ncd=\Zcd\Dcd$). 
The complex graph incidence matrix $\DD$ determines both the flow and
potential of closed systems, but in open systems the flow is
determined by $\Dcd$ and the potential by $\DD$.
As, in general $\Dcd \neq \DD$, the digraph for the flow of open
systems is not the same as the digraph for potentials. In particular,
with reference to Figure \ref{fig:transporter_graphs}, the flow and
potential digraphs for open systems may be structurally different.

The combination of the explicit energy-compliance feature of the bond
graph modelling approach with the generic results of the graph-theory
based chemical reaction network approach will, it is hoped, lead to
new results and methods for the analysis and synthesis of biomolecular
systems.

\section{Acknowledgements}
Peter Gawthrop would like to thank the Melbourne School of Engineering
for its support via a Professorial Fellowship.
This research was in part conducted and funded by the Australian
Research Council Centre of Excellence in Convergent Bio-Nano Science
and Technology (project number CE140100036).
The authors would like to thank Ivo Siekmann for alerting them to
references \citep{SchRaoJay13,SchRaoJay15,SchRaoJay16} and the
anonymous reviewers for helpful comments on the manuscript.


\end{document}